\documentclass[aps,prl,twocolumn,groupedaddress,showpacs,showkeys]{revtex4-1}
\usepackage{color}
\usepackage{amsmath}
\usepackage{amsfonts}
\usepackage{graphicx}
\usepackage{subfigure}

\ifx\pdftexversion\undefined
\usepackage[dvips]{hyperref}
\else
\usepackage{hyperref}
\fi
\hypersetup{
  colorlinks = true, linkcolor = magenta
}
\definecolor{darkgreen}{rgb}{0,0.4,0}
\renewcommand{\phi}{\varphi}

\renewcommand{\theta}{\vartheta}

\begin{document}

\title{Classification of Mixed State Topology in One Dimension}

\author{Evert~P.~L.~van\ Nieuwenburg}
\author{Sebastian~D.\ Huber}
\affiliation{Institute for Theoretical Physics, ETH Zurich, 8093 Z{\"u}rich, Switzerland}

\date{\today}

%TC:ignore 
\begin{abstract} 
We show how to generalize the concepts of identifying and classifying symmetry protected topological phases in 1D to the case of an arbitrary mixed state. The pure state concepts are reviewed using a concrete spin-$1$ model. For the mixed state setup we demonstrate our findings numerically using matrix product state algorithms. Starting from the ground state and applying various types of noise sources we find a transient regime where the system is driven out of equilibrium while retaining its topological properties.
\end{abstract}

%TC:endignore 
\maketitle
 
{\em Introduction --} Since the advent of topological insulators the study of  phases that are not amenable to a treatment by the Ginzburg-Landau scheme has seen a tremendous revival. In particular, a classification of ground-states that are separated by an {\em energy gap} from the lowest lying excitations and  {\em do not break any symmetry} of the Hamiltonian give rise to the notion of topology: Are there distinct classes of such states that cannot be transformed into each other without closing the excitation gap? Prime examples are the fractional quantum Hall states \cite{Wen95} or the ground states of certain spin systems \cite{Yan11}. Due to their remarkable properties, like the existence of non-abelian quasi-particles, these states are sought after both in classical solid state materials as well as in ``quantum engineered'' systems like cold atoms, arrays of superconducting qbits, coupled non-linear laser-cavities, etc. The flexibility in adjusting system parameters in these engineered systems often comes at the price of a poor coupling to a thermal bath. Hence, these systems are often driven out of equilibrium quite easily. This raises the question how the notion of a topological {\em state} is carried over to a classification of {\em mixed density matrices} \cite{Cas07a, Has11a, Viy13a}. In this paper we study the evolution of topological ground states in the presence of non-equilibrium noise and propose a classification scheme for the corresponding mixed density matrices in one spatial dimension.

The identification of topological (ground) states at zero temperature is well understood~\cite{Kitaev09,Ryu10,Wen95}. The method of identifying the topological properties depends on the system at hand: For the case of insulating free fermions one turns to topological invariants \cite{Ryu10,Hasan10,Qi11}. For interacting systems general features such as edge states, ground state degeneracies and excitation statistics are indicators of topology~\cite{Che11c}. One dimensional (1D) systems are somewhat special as topologically non-trivial phases can only emerge in the presence of symmetries~\cite{Che11b}. Projective representations  of the respective symmetry groups (PSG) provide a powerful tool to classify these 1D phases~\cite{Che11a,Pol12a}. Depending on the type of investigation, analytical, numerical, or experimental, a PSG analysis might not always be readily available. Before we embark on the task of classifying mixed density matrices we mention less definitive, but potentially better accessible, methods of describing topological properties of gapped 1D phases.

Entanglement plays a key role in the classification of topological states~\cite{Che11b}. For classical systems, correlation functions can be used to identify phases and phase transitions. It is only natural to expect that  ``quantum correlations'', i.e. entanglement, shows features of quantum phase transitions. By far the most studied type of entanglement is the one between two spatial sub-systems: One divides the system into two parts labeled $A$ and $B$, and considers the reduced density matrix of either of the two $\rho_A = \text{tr}_B \rho$. The eigenvalues of $\rho_A$ form the entanglement spectrum (ES)~\cite{Li08a}, and from their sum one can extract the so called entanglement entropy. Through the properties of both the ES and the entanglement entropy one can obtain information about the topological nature of the system~\cite{Lae13a}. We now address the issue of how one can transfer these concept to the description of mixed density matrices?

Let us turn back to the question of topological states in open systems. In order to assess the stability of a topological phase, a clear feature has to be found that uniquely identifies whether the state is in a trivial or non-trivial phase. Attempts at extending the topological invariants to open systems exist \cite{Viy13b, Avron11, Avron12}. These methods are restricted to specific types of master equations, however, and a general tool for the identification of open system topology is still missing. 

The main result of this paper is the introduction of a classification scheme for mixed density matrices of gapped 1D systems. It extends the well known PSG analysis for ground states~\cite{Che11a,Pol12a}. We also discuss how a generalization of the ES can be efficiently calculated and related to the aforementioned PSG analysis. We develop our ideas on the example of a concrete model, the antiferromagnetic spin-1 chain. Before we introduce our method and results we provide a detailed overview on the tools that have been introduced for the pure state case, which we then extend to the open system setting.

{\em Review --} In this section we introduce the concepts necessary for this paper. We develop all ideas on the basis of a concrete model, highlighting which features are model specific and which are general. We consider the antiferromagnetic spin-$1$ Heisenberg chain:
\begin{equation}
H = J\sum_i \mathbf{S}_i \cdot \mathbf{S}_{i+1} -
 B\sum_i S^x_i - D\sum_i \left(S^z_i\right)^2.
\label{eq:model}
\end{equation}
The $\mathbf{S}$ operators represent $S=1$ degrees of freedom, $B$ sets the strength of an external field and $D$ is an on-site anisotropy. For an extended region in the $B$--$D$ phase diagram around the Heisenberg point ($B = D = 0$), this model is in a symmetry-protected topological phase called the Haldane phase~\cite{Gu09a}. This Haldane phase region also includes the Affleck-Kennedy-Lieb-Tasaki model~\cite{Affleck87,Affleck88}.

The symmetries of this model responsible for the protection of Haldane phase are translation, time reversal and inversion symmetry~\cite{Gu09a}. At $B=0$ the model also has a $\mathbb{Z}_2 \times \mathbb{Z}_2$ symmetry (spin rotations of $\pi$ around two orthogonal axes), allowing for the definition of a non-local order parameter called string order \cite{Nijs89}
\begin{equation}
		\langle O^{x,z} \rangle = \lim_{|i-j|\to\infty} \langle \psi | \mathbf{S}_i^{x,z} e^{-i\pi \sum_{i<k<j} \mathbf{S}_k^{x,z}} \mathbf{S}_j^{x,z} | \psi \rangle.
\label{eq:StringOrder}
\end{equation}
The string order parameter in $z$ detects a dilute antiferromagnetic ordering of the spins in the $z$-direction, where the quantum numbers $m^z$ alternate perfectly between $+1$ and $-1$ after stripping out the $m^z = 0$ values (and similarly for the string order in $x$). For our model, a non-vanishing string order identifies the Haldane phase. For the cases of other symmetries than $\mathbb{Z}_2 \times \mathbb{Z}_2$, a generalized string order can be defined~\cite{Per08a}. Unfortunately, string order is restricted to 1D systems. Moreover it and cannot be used to fully classify 1D phases~\cite{Pol12a}. 

For the full classifcation of gapped 1D phases, we turn to the projective symmetry group analysis. This is most easily explained using the formalism of matrix product states (MPS). In an MPS, the coefficients of a general wavefuction, given in a local basis $|i\rangle$, are written in terms of a product of matrices:
\begin{equation}
	|\psi\rangle = \sum_{ij\ldots} 
	\text{Tr}\left(A_i^{[1]} A_j^{[2]}\cdots\right) |i\,j\cdots\rangle.
\label{eq:generalMPS} \end{equation}
For every site $n$, a set of $\chi \times \chi$ matrices $A^{[n]}_i$ is introduced (one for each possible basis state). In general, the bond dimension $\chi$ can be chosen such that the full wavefunction is exactly represented. For the ground state of 1D gapped Hamiltonians, such as Eq.~\ref{eq:model}, an accurate representation only requires a $\chi$ that grows algebraically with the number of sites. Hence, Eq (\ref{eq:generalMPS}) is a very efficient representation of such states.

We are now in the position to turn our attention to the PSG analysis. A Hamiltonian may be invariant under a certain set of symmetries $\mathcal{G}$. Every element $g \in \mathcal{G}$ has a representation on the wavefunctions, and in particular one may find a representation (a $\chi \times \chi$ matrix $U_g$) on the $A$ matrices:
\begin{equation}
	\tilde{A}_i^{[n]} = e^{i\theta(g)} U_g^\dagger \, A_i^{[n]} \, U_g.
\label{eq:PSGrep}
\end{equation}
As $U_g$ and $U_g^\dag$ always appear together in the physical state, the phase $\theta(g)$ does not affect the problem and renders the representation {\em projective}, i.e., one requires only $U_gU_{g'}=\exp[i\phi(g,g')]U_{gg'}$. It has been shown, that the freedom to assign the phases $\phi(g,g')$ gives rise to equivalence classes of representations that can be used to classify 1D SPT phases \cite{Che11b,Che11a,Pol12b} (also see the Appendices). For example for the $\mathbb Z_2 \times \mathbb Z_2$ symmetry this leads to classes where the matrices $U_g$ and $U_{g'}$ either anti-commute or commute. The determination of the $U_g$ matrices can be done efficiently using MPS algorithms see Ref~\cite{Pol12a} and Appendices A and B. Besides a simple platform for the study of PSGs, matrix product states have more advantages.

Within an MPS formulation, the entanglement spectrum is obtained for free.  The $A^{[n]}_i$ matrices in Eq.~\ref{eq:generalMPS} may be decomposed (via a singular value decomposition) into $\Gamma^{[n]}_i \lambda^{[n]}$, where the $\lambda$ matrices are diagonal matrices whose entries $\lambda_\alpha$ are related to the entanglement spectrum via $-2\ln \lambda_\alpha$.

The entanglement spectrum can also be used as a tool for studying topoplogical phases.  Since its introduction~\cite{Li08a}, its relation to topology has been extensively studied.  
In our case of the 1D SPT phases, the feature in the ES of a non-trivial state is that all the values  $\lambda_\alpha$ form pairs of even multiplicity~\cite{Pol10a}. This is fundamentally due to the projective symmetry representation matrices $U_g$ being antisymmetric~\cite{Pol10a}. Hence, it is indicating the PSG. However, a degenerate ES is not in one to one correspondence with the topology of the state. Hence, the ES is an easily available but not strictly conclusive tool to investigate gapped symmetric phases. For example, a state symmetric under both $\mathbb Z_2 \times \mathbb Z_2$ and inversion, could be trivial with respect to one but not the other PSG, but both would induce the same degeneracy in the ES.

{\em Method --} The MPSs introduced above, allow only for the simulation of pure state wavefunctions. The addition of an environment in the description of the system must be done on the level of the density matrix $\rho$. The evolution of $\rho$ is governed by a master equation:
\begin{equation}
	\dot{\rho}(t) \!=\! \mathcal{L}[\rho]\!=\! -\frac{i}{\hbar}[H, \rho]\! + \gamma
\sum_i
\!\left[L_i^\dagger L_i\rho -\! \frac{1}{2}\!\left\{ \rho, L_i L_i^\dagger \right\}\!\right],
\label{eq:Lindblad}
\end{equation}
in which the jump operator $L$ represents the system part of the coupling to the environment. We shall consider different jump operators in our discussion of the results. We first turn to the simulation of the density matrix.

The MPS formalism can be extended to simulating the density matrix via the superoperator approach~\cite{Zwo04a,Oru08a}.  The density matrix is expanded in a set of local basis matrices ${\sigma_i}$, after which the coefficients in this expansion are written as a product over matrices:
\begin{align}
	\rho &= \sum_{ijk\ldots} \rho_{ijk\ldots} \mathbf{\sigma}_i \mathbf{\sigma}_j \mathbf{\sigma}_k \cdots,\\
	\rho_{ijk\ldots} &= \text{Tr}\left(A^{[1]\sharp}_i A^{[2]\sharp}_j A^{[3]\sharp}_k \cdots \right).
\label{eq:MPS}
\end{align}
The density matrix $\rho$ is then interpreted as a vector $|\rho\rangle_\sharp$, on which superoperators act as matrices.  A superoperator $T_A$ that represents multiplication by a matrix $A$, i.e. $T_A[\rho] = A\rho$, is turned into a matrix $T_A^\sharp$ acting on the vectorized density matrix as \[ T_A^\sharp|\rho\rangle_\sharp = | T_A[\rho] \rangle_\sharp = | A\rho \rangle_\sharp. \] In particular the master equation in Eq.~\ref{eq:Lindblad} turns into $|\dot{\rho}\rangle_\sharp = \mathcal{L}_\sharp|\rho\rangle_\sharp$, and becomes formally equivalent to the Schr\"odinger equation.  This allows us to simulate the density matrix MPS using the same algorithms as for the pure state MPS~\cite{Zwo04a,Verstraete04,Vid07a}.

Using the superoperator approach we are able to go through the program outlined on the pure-state problem: (i) We can calculate expectation values and correlation functions (such as string order); (ii) we can determine the action of projective symmetries on the MPS to obtain the $U_g$ matrices; (iii) and we can calculate the superentanglement spectrum.

The superentanglement spectrum (SES), $\lambda_\sharp$, is obtained from the $A^{[n]\sharp}$ matrices via a decomposition into $\Gamma_\sharp \lambda^{[n]}_\sharp$ in the same way es the ES is obtained in the pure state case.  Let us analyze what we can learn from the SES.  If $|\rho\rangle_\sharp$ describes a pure state, it is straightforward to show that $\lambda_\sharp = \lambda \otimes \lambda$.  For a general mixed state, the relation between the two becomes more involved. Fortunately, the degeneracy (pairs of even mutliplicity) of the ES is a property that comes from a symmetry of the state, and therefore persists also for the SES.  This allows us to directly interpret the degeneracy of the SES as coming from a topologically non-trivial state. In the following sections, we demonstrate the use of the superoperator approach on the introduced model (Eq.~\ref{eq:model}).  

{\em Results --} As a first step, we calculate the thermal density matrix using the superoperator approach \cite{Oru08a}. Starting from an infinite temperature state, $\rho \sim \openone$, any finite temperature state can be reached by applying the following evolution:
\begin{equation}
	|\rho_\beta\rangle_\sharp = |e^{-\beta H}\rangle_\sharp = e^{-\beta T_\sharp}|1\rangle_\sharp.
\label{eq:Cooling}
\end{equation}
Here $T_\sharp$ is a superoperator corresponding to $T[A] = \frac{1}{2}(AH + HA)$ with $H$ the Hamiltonian in Eq.~\ref{eq:model} \cite{Oru08a}.

\begin{figure}[t]
\includegraphics[width=\columnwidth]{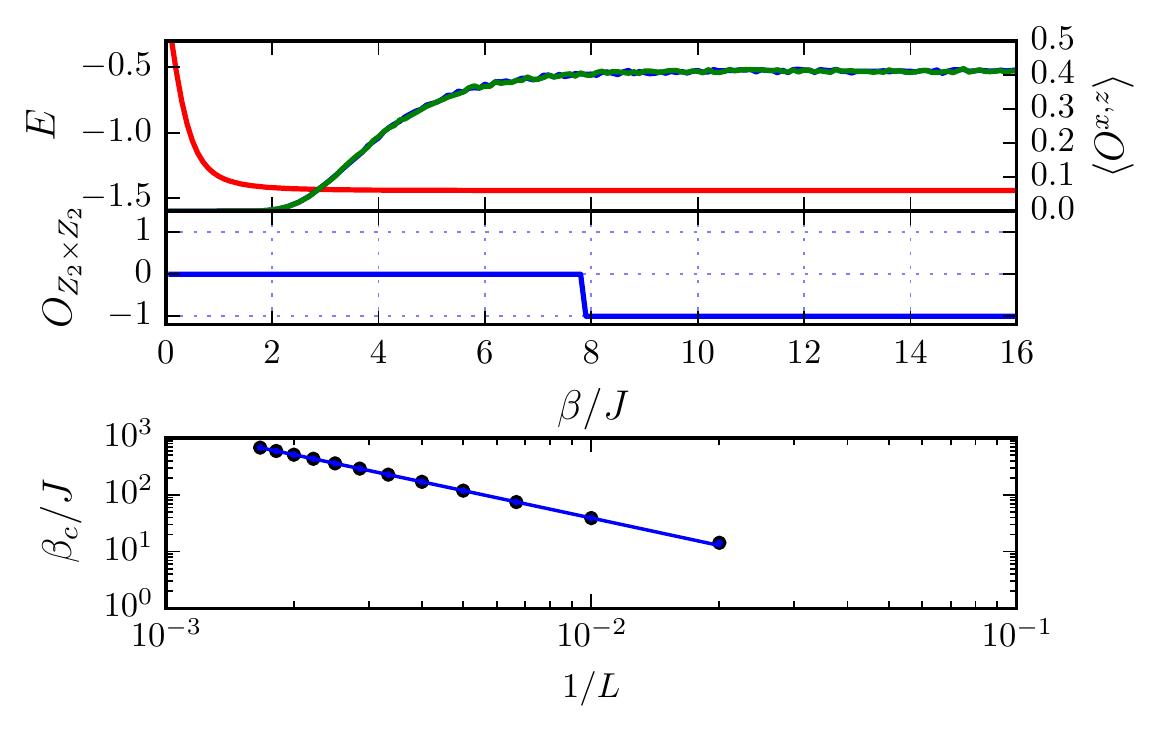}
\caption{
The top panel shows the convergence of energy and string order while cooling. The $\mathcal{O}_{Z_2\times Z_2}$ trace shows the state going from non-symmetric to topological. The bottom panel shows that for finite systems of size $L$	there is a finite temperature $\beta_c$ at which the string order correlation length diverges, i.e., string order only develops at $T=0$.
}
\label{fig:Cooling}
\end{figure}

The results for the approach of the energy towards the ground  state at the Heisenberg point ($B = D = 0$) are presented in the top panel of Fig.~\ref{fig:Cooling}. The figure also shows the emergence of the string order parameter $\langle O^{x,z} \rangle$ and the result of the PSG analysis. We show $\mathcal{O}_{Z_2 \times Z_2}$, computed from the represenation matrices $U_g$ of the $\mathbb{Z}_2\times\mathbb{Z}_2$ symmetry on the MPS (Appendices). This parameter can take the values $0$ if the state is not symmetric, or $\pm 1$ for a symmetric and trivial or non-trivial state, respectively. These measurements were performed using an MPS of maximal bond dimension $\chi=100$, keeping the truncation error below $10^{-8}$ at all times.
The top panel of Fig.~\ref{fig:Cooling} shows finite string order developing at small but non-zero temperatures, whereas string order is expected only to develop for zero temperature and infinite systems. This effect is due to the finite size of the system (see bottom panel and figure caption).

\begin{figure}[t!]
	\includegraphics[width=\columnwidth]{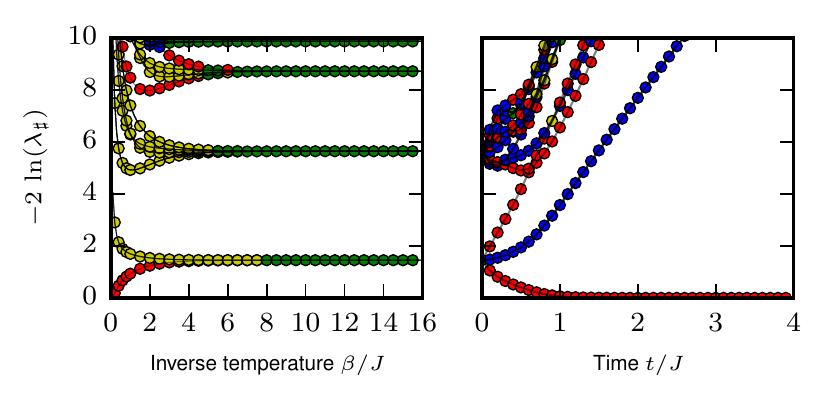}
	\caption{
(Left) The SES as a function of inverse temperature (only the $42$ lowest values are shown). When the temperature is lowered, the expected pairs with multiplicities being multiples of four develops. For the color code, see Tab.~\ref{tab:colors}
(Right) Time evolution of the ground state at the Heisenberg point $(B=D=0)$ with noise strength $\gamma = 0.5$ for a local jump operator $S^-$ pumping the spins into the minimal $S^z$ eigenstate. Only the lowest 20 values are shown.
}
\label{fig:CoolingSES}
\end{figure}

We now turn to the discussion of the SES.  The ground state in the Haldane phase has an ES in which the multiplicity of the lowest pair is two.  In Fig.~\ref{fig:CoolingSES} the SES is seen to develop a multiplicity of at least four. The fourfold multiplicity instead of two is a direct consequence of the $\lambda_\sharp$ being related to the $\lambda$ via a tensorproduct.
\begin{table}[t]
\begin{ruledtabular}
\begin{tabular}{llll}
\multicolumn{2}{l}{degeneracy ($n \in \mathbb N$, $n>0$)} & significance& color \\
\hline
$1$ &$\!\!\!\!\!\bullet\circ\circ\circ\circ\circ\circ\circ\circ\circ\dots$   & trivial & {\color{red}red}\\
$2n-1$ & $\!\!\!\!\!\circ\circ\bullet\circ\bullet\circ\bullet\circ\bullet\circ\dots$  & trivial & {\color{yellow}yellow} \\
$2(2n-1)$ & $\!\!\!\!\!\circ\bullet\circ\circ\circ\bullet\circ\circ\circ\bullet\dots$& non-trivial/mixed &  {\color{blue}blue} \\
$4n$ &$\!\!\!\!\!\circ\circ\circ\bullet\circ\circ\circ\bullet\circ\circ\dots$  & non-trivial/pure& {\color{darkgreen}green} \\
\end{tabular}
\end{ruledtabular}
\caption{
Color code for SES plots.
}
\label{tab:colors}
\end{table}

So far we have shown the development of the topologically non-trivial Haldane phase via the emergence of string order, the PSG, and the super entanglement spectrum. Starting from the ground state obtained from the cooling, the evolution of the state governed by Eq.~\ref{eq:Lindblad} can be simulated for various types of jump operator $L$.

We intoduce three different jump operators and explain where they find an application. First, 
\begin{equation}
L_i = S_i^-
\end{equation}
represents a case where the spins slowly relax towards the minimum $m_z=-1$. For an engineered system with $B\neq 0$ this corresponds to a decay of the qdit making up the local spin-1. We then consider the influence of a fluctuating external field, 
\begin{equation}
\label{eqn:fmf}
L_i = S^z_i.
\end{equation}
Another potentially important term is a fluctuating Ising coupling 
\begin{equation}
L_i = S^z_iS^z_{i+1}.
\end{equation}
For an implementation of a spin-chain with trapped ions this correspond to fluctuations in the control gates responsible for the couplings \cite{Kim10}.
  
For $L_i = S_i^-$, we expect for the steady state a trivial product state. This is confirmed by the single non-degenerate value in the SES in Fig.~\ref{fig:CoolingSES}. The immediate splitting of the spectrum indicates that this particular type of pump destroys the symmetries protecting the topological phase. Varying the noise strength $\gamma$ simply sets the timescale for when the steady state is reached.

\begin{figure}[t]
\includegraphics[width=\columnwidth]{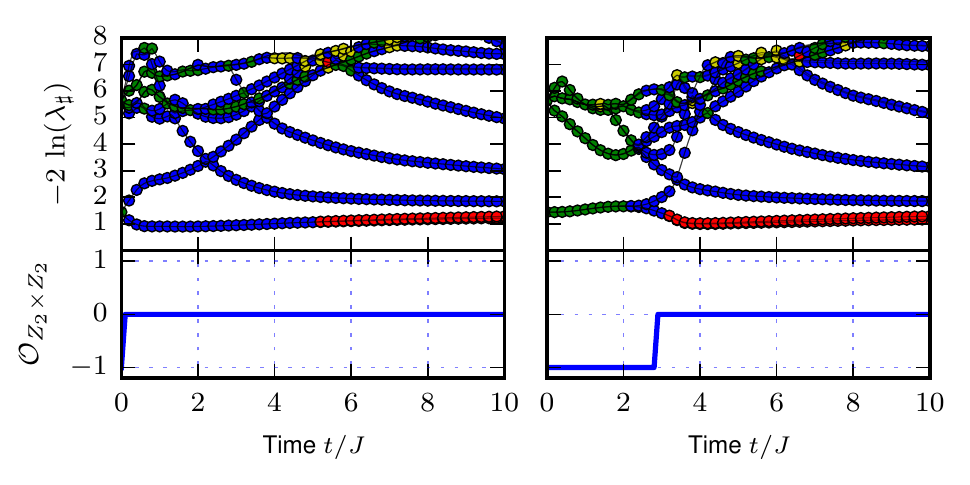}
\caption{
Superentanglement spectra for $L=S^z$ (left) and $L=S^zS^z$ (right), starting from $B=D=0$ with $\gamma = 0.5$. Only the lowest 20 values are shown.
}
\label{fig:NoiseSES}
\end{figure}

We now turn to the fluctuating magnetic field (\ref{eqn:fmf}). Switching on the noise governed by $L_i = S_i^z$ breaks the $\mathbb{Z}_2 \times \mathbb{Z}_2$ symmetry. This is reflected in the PSG parameter $\mathcal{O}_{\mathbb{Z}_2 \times \mathbb{Z}_2}$ in Fig.~\ref{fig:NoiseSES}. The figure also shows that the degeneracy in the SES is present nevertheless. The loss of this symmetry is therefore not enough for the system to lose its topological nature. For longer times however, the superentanglement spectrum splits and the symmetry protected topological state is lost.
A two-fold degeneracy in the super entanglement spectrum excludes the relation $\lambda_\sharp = \lambda \otimes \lambda$, indicating that the state is not a pure state. Hence a state obtained by stopping the time evolution with Lindblad operators at this point would be a topological mixed state. We have checked the convergence of our simulations by performing a scaling analysis in the bond dimension of our MPS (Appendix C).

The fluctuating Ising coupling, $L_i = S_i^zS_{i+1}^z$, preserves the $\mathcal{O}_{\mathbb{Z}_2 \times \mathbb{Z}_2}$ symmetry in the transient regime, cf. Fig.~\ref{fig:NoiseSES}. The symmetry is lost only later when the super entanglement spectrum loses its degeneracy completely.  Also in this case the degeneracy in the spectrum splits eventually. But before this happens, the transient regime hosts mixed states for which the super entanglement spectrum consists exclusively of pairs of even multiplicity.

{\em Conclusion --} The manifestations of topology in out-of-equilibrium systems is an important issue, especially with respect to experimental implementations and realizations of quantum information protocols. We have considered a generic example of a 1D symmetry protected topological phase, and studied the degeneracy of the super entanglement spectrum. These degeneracies can be directly linked to the topological phase, and survive in the transient regime before reaching a steady-state. This shows that for suitable types of noise, the mixed state may retain a notion of the topological state.

%TC:ignore 
\begin{acknowledgments} 
We happily acknoweldge discussion with Frank Pollman. This research was supported by the Swiss National Science Foundation.
\end{acknowledgments}

\appendix
\section*{Appendix}
These appendices provides details to statements in the main text.
Its main purpose is to describe the analysis of the projective symmetry group, and how to obtain such information in a numerically
efficient way. For coherence and consistency, we begin by describing how expectation values are measured on the pure and mixed 
state matrix product states. This allows for the introduction of a central concept, the `transfer matrix', which is key in the
understanding of the computation of the projective symmetry group matrices.

\section{A. Measuring expectation values \label{app:A}}
Compared to the pure state algorithms, there are minor differences in the way observables are measured on the mixed state. This section serves to illuminate them, by first reviewing how expectation values are measured in the pure state case. To that end, we first introduce (the standard) diagramattical represenation of matrix product states (MPS).

The matrices $A^{[n]}_i$ in the expansion of the state as an MPS are tensors with three indices. The $i$ index is referred to as the ‘physical leg’, since it represents the physical basis states. For each $i$ the $A_i^{[n]}$ is a matrix, so that $A$ in full notation we should include the indices $\alpha$ and $\beta$ as (dropping the site index $[n]$ for the moment) $A_i^{\alpha\beta}$. This object can then be represented diagrammatically where the extruding legs represent the indices
\begin{equation}
A_{i}^{\alpha\beta} = \raisebox{-0.5\height}{\includegraphics{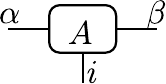}}.
\end{equation}
Contracting the legs of two $A$-matrices implies a summation over them
\begin{equation}
\sum_{\gamma} A_i^{\alpha\gamma}A_{j}^{\gamma\beta} = 
\raisebox{-0.5\height}{\includegraphics{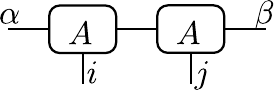}}.
\end{equation}
Similarly for the multiplication of $A$ with a physical operator $\mathcal O$
\begin{equation}
\sum_j \mathcal O_{ij}A_j^{\alpha\beta}   = 
\raisebox{-0.5\height}{\includegraphics{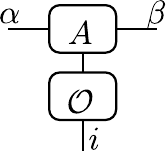}}.
\end{equation}
This allows us to represent the expectation value of an operator $\mathcal O$ by ``sandwiching'' it between the matrix product state represenations of $|\psi\rangle$ and $\langle \psi |$. 
If we consider a three-site chain as an illustration, we can express the measurement of a local operator on the middle site as:
\begin{equation}
\langle \psi | \mathcal{O}^{[2]} | \psi \rangle = \langle \psi | \openone \otimes \mathcal{O}^{[2]} \otimes \openone | \psi \rangle =
\raisebox{-0.5\height}{\includegraphics{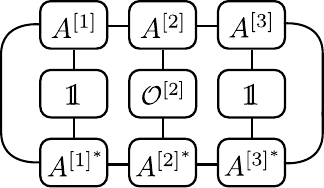}}.
\label{eq:fullsandwich}
\end{equation}
If the MPS is in canonical form (which will be explained below), the measurement of a local operator $\mathcal{O}_n$ on site $n$ can be computed from only the matrices representing that site. The previous expectation value is thus equivalent to
\begin{equation}
\langle \psi | \mathcal{O}^{[2]} | \psi \rangle = 
\raisebox{-0.5\height}{\includegraphics{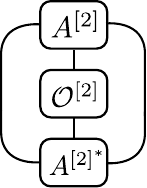}}.
\label{eq:localsandwich}
\end{equation}
The transfer ``matrix'' is defined as 
\begin{equation}
T_{\alpha\alpha',\beta\beta'}=
\raisebox{-0.5\height}{\includegraphics{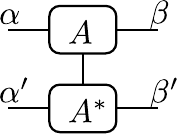}},
\end{equation}
and plays a central role in MPS based algorithms.  
If the MPS is in canonical form (the following may be taken as a definition of canonical form), the $A$ matrices are chosen such that $T_{\alpha\alpha',\beta\beta'}$ has a (unique) right and left dominant eigenmatrix $X$  with an eigenvalue $\eta$ with unit modulus $|\eta|=1$.
As an illustration for the above introduced graphical representation of tensors we write this as
\begin{equation}
\label{eqn:dominant}
\raisebox{-0.4\height}{\includegraphics{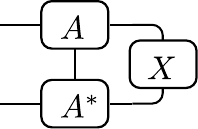}} = 
\eta\;
\raisebox{-0.38	\height}{\includegraphics{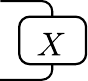}},
\;\; \mbox{with}\;\
\raisebox{-0.38	\height}{\includegraphics{X}}=\openone = \raisebox{-0.38	\height}{\includegraphics{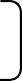}},\end{equation}
where in a further simplification we dropped the labels on the open legs. It is now clear why Eq.~\ref{eq:fullsandwich} reduces to Eq.~\ref{eq:localsandwich} if the MPS is in canonical form.

For the mixed state case, let us consider the case of a spin-$1/2$ system. On every site, we introduce a basis for $2\times 2$ matrices: $\sigma_{i=0,1,2,3} = \{\openone, \sigma_x, \sigma_y, \sigma_z\}$. It is imporantant that all of the matrices except for $\sigma_0$ are traceless, $\text{tr} \sigma_{i\neq 0} = 0$.
The expectation value of a local operator is written as $\langle \mathcal O \rangle = \text{Tr}( \rho \mathcal O )$. For the pure state case, the cyclic property of the trace led to this being equivalent to the sandwiching of the operator between two MPS copies. 

In the mixed state case, the measurement of the expectation value can be captured by a superoperator $T_{\mathcal O}[\rho] = \text{Tr}( \rho \mathcal O)$. One may compute the matrix elements of $T_{\mathcal O}^\sharp$ via $T^\sharp_{i,j} = {}_\sharp\langle \sigma_i | T[\sigma_j] \rangle_\sharp$, where the inner product between two vectorized $2\times 2$ matrices is given by ${}_\sharp \langle A | B \rangle_\sharp = \frac{1}{2}\text{Tr}(A^\dagger B)$. 
By virtue of $\sigma_0$ being the only basis matrix with a non-vanishing trace, the resulting expression for the expectation value can be diagramatticaly represented by
\begin{equation}
\text{Tr}( \rho \mathcal{O}^{[1]}\mathcal{O}^{[2]} ) = 
\raisebox{-0.5\height}{\includegraphics{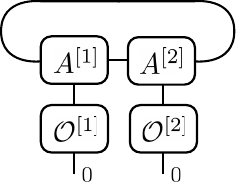}},
\end{equation}
where the $0$'s represent selecting the only the $i = 0$ coefficient on each matrix, and the trace is obtained by closing the bonds up top.

\section{B. Determining the $U_g$ matrices \label{app:B}}
The computation of the matrix representation of the projective symmetries can be efficiently done on the MPS of the density matrix. 

Using the diagramattical representation introduced above, we are now in the position to find the representation matrices $U_g$. Let us restrict the exposition here to symmetries that do not mix different sites, e.g., spin rotations. The generalization to translation or inversion symmetries, etc, is then straightforward. Each element of the symmetry group $g \in \mathcal G$ as a natural representation $R_g$ on the physical Hilbert space. We are looking for a represantion on the bond degrees of freedom
\begin{equation}
\raisebox{-0.5\height}{\includegraphics{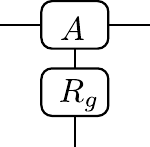}}
= e^{i\phi}\;
\raisebox{-0.5\height}{\includegraphics{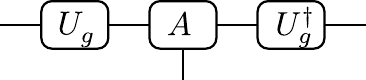}}.
\end{equation}
We now show that the generalized transfer matrix 
\begin{equation}
\label{eqn:generalized}
\raisebox{-0.5\height}{\includegraphics{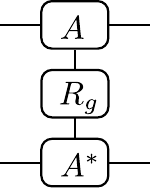}}
\end{equation}
has as its dominant eigenmatrix $U_g$ with eigenvalue $\lambda e^{2i\phi}$. We first use the definition of the $U_g$ matrices
\begin{align}
\raisebox{-0.5\height}{\includegraphics{generalizedtransfer}} 
= e^{2i\phi}\;
\raisebox{-0.5\height}{\includegraphics{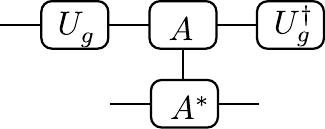}}
\end{align}
Dropping the phase factor $e^{2i\phi}$ we now continue
\begin{multline}
\label{eqn:proof}
\raisebox{-0.5\height}{\includegraphics{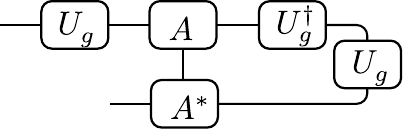}}=\\
\raisebox{-0.5\height}{\includegraphics{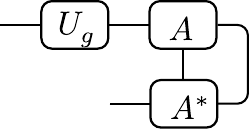}}
=\eta \raisebox{-0.4\height}{\includegraphics{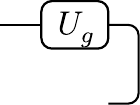}}
=\eta \raisebox{-0.5\height}{\includegraphics{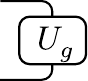}}.
\end{multline}
In the first step we made use of the unitarity of $U_g$ and in the second step we used (\ref{eqn:dominant}). With this we proved that the generalized transfer matrix has a dominant eigenmatrix $U_g$. 

Using Eq. (\ref{eqn:proof}) we have a simple recipe at hand. (i) Find the MPS of the density matrix, i.e., the $A$-matrices. (ii) Construct the generalized transfer matrix ({\ref{eqn:generalized}). (iii) Find the dominant eigenmatrix, which is the sought after $U_g$. Note that if the
dominant eigenvalue is not of unit modulus, the state was evidently not invariant under the symmetry operation and the $U_g$ matrix can be disregarded.

With the $U_g$ matrices we are now able to determine whether the state is in a topologically trivial or non-trivial phase. We will now consider one such calculation in more detail for the case of the $\mathbb{Z}_2\times\mathbb{Z}_2$ symmetry. 
If a state has this symmetry, $180$-degree rotations around the $x$ and $z$ axes leave it invariant. For each of these two operations, $\mathcal{R}_x$ and $\mathcal{R}_z$, we can obtain the matrix representations $U_x$ and $U_z$ of how they act on the MPS $A$ matrices. 
Rotating by $180$-degrees twice will give back the same state up to a possible phase, so that $U_x^2 = e^{i\phi}\openone$ and similar for $U_z$. 
One may choose to absorb this phase in the definition of the $U_x$ matrix, gauging it away. 
But this gauge freedom does not exist for all of the elements of the symmetry group $\mathbb{Z}_2 \times \mathbb{Z}_2$.
In particular, one may consider the operations $\mathcal{R}_x\mathcal{R}_z$ and $\mathcal{R}_z\mathcal{R}_x$. The representations of these operations have a relative phase of $\pm 1$ that cannot be gauged away. In other words, $U_xU_z = \pm U_zU_x$ and hence $U_x$ and $U_z$ either commute or anti-commute. 
This can be captured in a number by computing
\[ \mathcal{O}_{\mathbb{Z}_2\times\mathbb{Z}_2} = \frac{1}{\chi}\text{Tr}( U_x U_z U_x^\dagger U_z^\dagger ), \]
which evaluates to $\pm 1$ if the matrices commute or anti-commute respectively. If the state was not symmetric, this analysis might still produce a $\pm 1$, but one manually sets its value to zero in such a case.

\section{C. Scaling of bond dimension \label{app:C}}
\begin{figure}[htbp!]
    \includegraphics[width=\columnwidth]{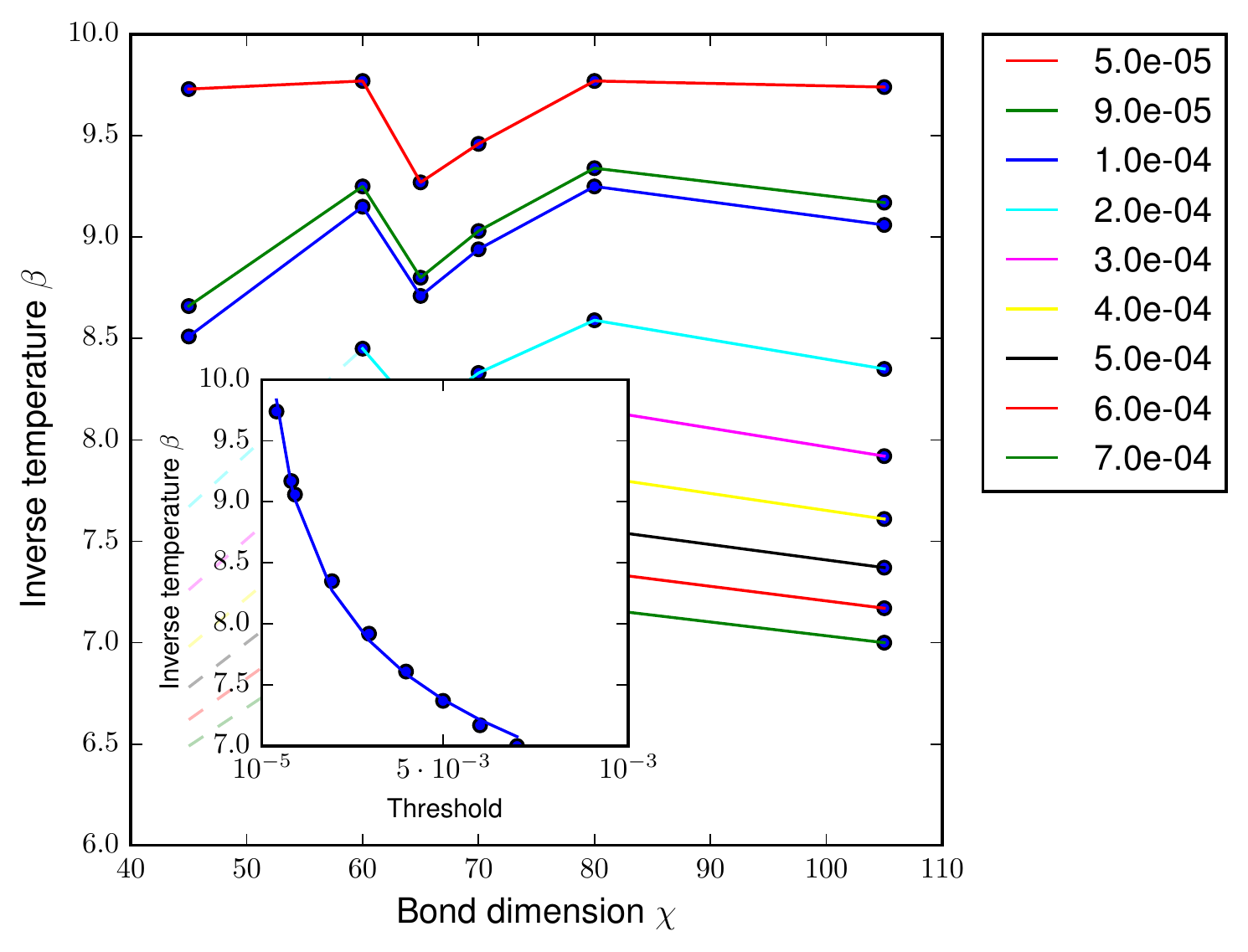}
    \caption{
The temperature at which a degeneracy of the lowest two super entanglement values is reached (the legend shows the degeneracy thresholds). 
For bond dimensions of about $60$ and higher, the transition temperature shows little variance. 
The inset shows a cut of the degeneracy at $\chi = 105$, and an inverse power law fit indicating that full degeneracy is reached only at $\beta = \infty$.
\label{fig:scaling}}
\includegraphics[width=\columnwidth]{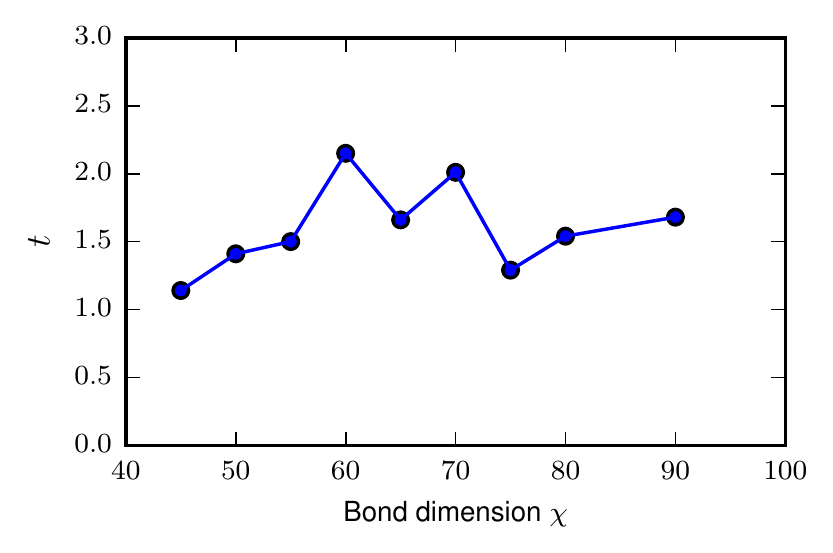}
    \caption{
The time point at which the four-fold degeneracy splits into a two times two-fold degenerate pair for increasing bond dimension $\chi$.
\label{fig:noisescaling}}
\end{figure}

In order to check the convergence of the MPS method, a comparison between different bond dimensions $\chi$ is crucial. 
We perform a scaling analysis of the point where the state becomes topologically non-trivial in the cooling simulations. 
For a fixed degeneracy threshold, we track the (inverse) temperature at which the lowest two super entanglement values become degenerate as a function of bond dimension $\chi$.
The results for a (small) degeneracy threshold of about $10^{-4}$ are shown in Fig.~\ref{fig:scaling}, indicating that from bond dimensions about $60$ and higher little variance is found in the transition temperature. Decreasing the threshold shows that full degeneracy is only reached for zero temperature.
A similar analysis can be performed on the noise traces (see Fig.~\ref{fig:noisescaling}). A 'landmark' point can be set at the point where the lower four-fold degenerate pair splits into two two-fold degenerate pairs. The exact point at which this happens for various bond dimensions differs, but does not seem to indicate trend towards immediate splitting upon increasing the bond dimension.


%TC:endignore 
\end{document}